\documentclass[twocolumn,showpacs,prc,floatfix]{revtex4}
\usepackage{graphicx}
\usepackage[dvips,usenames]{color}
\usepackage{amsmath,amssymb}
\newcommand{\bs}{\begin{sloppypar}} \newcommand{\es}{\end{sloppypar}}
\def\beq{\begin{eqnarray}} \def\eeq{\end{eqnarray}}
\def\beqstar{\begin{eqnarray*}} \def\eeqstar{\end{eqnarray*}}
\newcommand{\bal}{\begin{align}}
\newcommand{\eal}{\end{align}}
\newcommand{\beqe}{\begin{equation}} \newcommand{\eeqe}{\end{equation}}
\newcommand{\p}[1]{(\ref{#1})}

\begin {document}
\title{Spin polarized states in neutron matter at a strong
magnetic field
 }
\author{ A. A. Isayev}
\email{isayev@kipt.kharkov.ua}
 \affiliation{Kharkov Institute of
Physics and Technology, Academicheskaya Street 1,
 Kharkov, 61108, Ukraine
\\
Kharkov National University, Svobody Sq., 4, Kharkov, 61077,
Ukraine
 }
  \author{J. Yang}
 \email{jyang@ewha.ac.kr}
 \affiliation{Department  of Physics and the Institute for the Early Universe,
 \\
Ewha Womans University, Seoul 120-750, Korea
}
\begin{abstract}
Spin polarized states  in neutron matter at  strong magnetic
fields up to $10^{18}$~G are considered in the model with the
Skyrme effective interaction. By analyzing the self-consistent
equations at zero temperature, it is shown that a
thermodynamically stable branch of solutions for the spin
polarization parameter as a function of density corresponds to the
negative spin polarization  when the majority of neutron spins are
oriented opposite to the direction of the magnetic field. Besides,
beginning from some threshold density dependent on the magnetic
field strength the self-consistent equations have also two other
branches of solutions for the spin polarization parameter with the
positive spin polarization.
 The free energy corresponding to one of these branches turns out
 to be very close to that of the thermodynamically preferable branch.
  As a consequence, at a strong magnetic field, the
state with the positive spin polarization can be realized as a
metastable state at the high density region in neutron matter
which under decreasing density  at some threshold density changes
into a thermodynamically stable state with the negative spin
polarization.
\end{abstract}
\pacs{21.65.Cd, 26.60.-c, 97.60.Jd, 21.30.Fe} \keywords{Neutron
star models, magnetar, neutron matter, Skyrme interaction, strong
magnetic field, spin polarization} \maketitle

\section{Introduction}
Neutron stars observed in nature are magnetized objects with the
magnetic field strength at the surface in the range of
$10^{9}$-$10^{13}$~G~\cite{LGS}. For a special class of  neutron
stars such as soft gamma-ray  repeaters  and anomalous X-ray
pulsars, the field strength can be much larger and is estimated to
be  about $10^{14}$-$10^{15}$~G~\cite{TD}. These strongly
magnetized objects are called magnetars~\cite{DT} and comprise
about $10\%$ of the whole population of neutron stars~\cite{K}.
However, in the interior of a magnetar the magnetic field strength
may be even larger, reaching values of about
$10^{18}$~G~\cite{CBP,BPL}. The possibility of existence of such
ultrastrong magnetic fields is not yet excluded, because what we
can learn from the magnetar observations by their periods and
spin-down rates, or by hydrogen spectral lines is only their
surface fields. There is still no general consensus regarding the
mechanism to generate such strong magnetic fields of magnetars,
although different scenarios were suggested such as, e.g., a
turbulent dynamo amplification mechanism in a neutron star with
the rapidly rotating core at first moments after it goes
supernova~\cite{TD}, or the possibility of spontaneous spin
ordering in the dense quark core of a neutron star~\cite{ST}.

Under such circumstances, the issue of interest is the behavior of
 neutron star matter in a strong magnetic
field~\cite{CBP,BPL,CPL,PG}. In the recent study~\cite{PG},
neutron star matter was approximated by  pure neutron matter in
the model considerations with the effective Skyrme and Gogny
forces. It has been shown that the behavior of the spin
polarization of neutron matter in the high density region at a
strong magnetic field crucially depends on whether neutron matter
develops a spontaneous spin polarization (in the absence of a
magnetic field) at  several times  nuclear matter saturation
density as is usual for the Skyrme forces, or the appearance of a
spontaneous polarization is not allowed  at the relevant densities
(or delayed to much higher densities), as in the case with the
Gogny D1P force. In the former case, a ferromagnetic transition to
a totally spin polarized state occurs while in the latter case a
ferromagnetic transition is excluded at all relevant densities and
the spin polarization remains quite low even in the high density
region. Note that the issue of spontaneous appearance of spin
polarized states in neutron and nuclear matter is a controversial
one. On the one hand, the models with the Skyrme effective
nucleon-nucleon (NN) interaction predict the occurrence of
spontaneous spin instability in nuclear matter at densities in the
range from $\varrho_0$ to $4\varrho_0$ for different
parametrizations of the NN potential~\cite{R}-\cite{RPV}
($\varrho_0 = 0.16\,{\rm fm}^{-3}$ is the nuclear saturation
density). For the Gogny effective interaction, a ferromagnetic
transition in neutron matter occurs at densities larger than
$7\varrho_0$ for the D1P parametrization and is not allowed for
D1, D1S parametrizations~\cite{LVRP}. However, for the D1S Gogny
force an antiferromagnetic phase transition happens in symmetric
nuclear matter at the density $3.8\varrho_0$~\cite{IY2}. On the
other hand, for the models with the realistic NN interaction, no
sign of spontaneous spin instability  has been found so far at any
isospin asymmetry up to densities well above
$\varrho_0$~\cite{PGS}-\cite{BB}.

Here we study  thermodynamic properties of spin polarized neutron
matter at a  strong magnetic field in the model with the Skyrme
effective forces. As a framework for consideration, we choose a
Fermi liquid approach for the description of nuclear
matter~\cite{AKPY,AIP,IY3}. Proceeding from the minimum principle
for the thermodynamic potential, we get the self-consistent
equations for the spin order parameter and chemical potential of
neutrons. In the absence of a magnetic field, the self-consistent
equations have two degenerate branches of solutions for the spin
polarization parameter corresponding to the case, when the
majority of neutron spins are oriented along, or opposite to the
spin quantization axis (positive and negative spin polarization,
respectively). In the presence of a magnetic field, these branches
are modified differently. A thermodynamically stable branch
corresponds to the state with the majority of neutron spins
aligned opposite to the magnetic field. At a strong magnetic
filed, the branch corresponding to the positive spin polarization
splits into two branches with the positive spin polarization as
well. The last solutions were missed in the study of
Ref.~\cite{PG}. We perform a thermodynamic analysis based on the
comparison of the respective free energies and arrive at the
conclusion about the possibility of the formation of metastable
states in neutron matter with the majority of neutron spins
directed along the strong magnetic field. The appearance of such
metastable states can be possible due to the strong spin-dependent
medium correlations in neutron matter with the Skyrme forces at
high densities.

Note that we consider  thermodynamic properties of spin polarized
states in neutron  matter at a strong magnetic field up to the
high density region relevant for astrophysics. Nevertheless, we
take into account the nucleon degrees of freedom only, although
other degrees of freedom, such as pions, hyperons, kaons, or
quarks could be important at such high densities.

\section{Basic equations}
 The normal (nonsuperfluid) states of neutron matter are described
  by the normal distribution function of neutrons $f_{\kappa_1\kappa_2}=\mbox{Tr}\,\varrho
  a^+_{\kappa_2}a_{\kappa_1}$, where
$\kappa\equiv({\bf{p}},\sigma)$, ${\bf p}$ is momentum, $\sigma$
is the projection of spin on the third axis, and $\varrho$ is the
density matrix of the system~\cite{I,IY}. Further it will be
assumed that the third axis is directed along the external
magnetic field $\bf{H}$. The energy of the system is specified as
a functional of the distribution function $f$, $E=E(f)$, and
determines the single particle energy
 \begin{eqnarray}
\varepsilon_{\kappa_1\kappa_2}(f)=\frac{\partial E(f)}{\partial
f_{\kappa_2\kappa_1}}. \label{1} \end{eqnarray} The
self-consistent matrix equation for determining the distribution
function $f$ follows from the minimum condition of the
thermodynamic potential~\cite{AKPY,AIP} and is
  \begin{eqnarray}
 f=\left\{\mbox{exp}(Y_0\varepsilon+
Y_4)+1\right\}^{-1}\equiv
\left\{\mbox{exp}(Y_0\xi)+1\right\}^{-1}.\label{2}\end{eqnarray}
Here the quantities $\varepsilon$ and $Y_4$ are matrices in the
space of $\kappa$ variables, with
$Y_{4\kappa_1\kappa_2}=Y_{4}\delta_{\kappa_1\kappa_2}$, $Y_0=1/T$,
and $ Y_{4}=-\mu_0/T$  being
 the Lagrange multipliers, $\mu_0$ being the chemical
potential of  neutrons, and $T$  the temperature.

Given the possibility for alignment of  neutron spins along or
opposite to the magnetic field $\bf H$, the normal distribution
function of neutrons and single particle energy can be expanded in
the Pauli matrices $\sigma_i$ in spin
space
\begin{align} f({\bf p})&= f_{0}({\bf
p})\sigma_0+f_{3}({\bf p})\sigma_3,\label{7.2}\\
\varepsilon({\bf p})&= \varepsilon_{0}({\bf
p})\sigma_0+\varepsilon_{3}({\bf p})\sigma_3.
 \nonumber
\end{align}

Using Eqs.~\p{2} and \p{7.2}, one can express evidently the
distribution functions $f_{0},f_{3}$
 in
terms of the quantities $\varepsilon$: \begin{align}
f_{0}&=\frac{1}{2}\{n(\omega_{+})+n(\omega_{-}) \},\label{2.4}
 \\
f_{3}&=\frac{1}{2}\{n(\omega_{+})-n(\omega_{-})\}.\nonumber
 \end{align} Here $n(\omega)=\{\exp(Y_0\omega)+1\}^{-1}$ and
 \bal
\omega_{\pm}&=\xi_{0}\pm\xi_{3},\label{omega}\\
\xi_{0}&=\varepsilon_{0}-\mu_{0},\;
\xi_{3}=\varepsilon_{3}.\nonumber\end{align}

As follows from the structure of the distribution functions $f$,
the quantity $\omega_{\pm}$, being the exponent in the Fermi
distribution function $n$, plays the role of the quasiparticle
spectrum. The spectrum is twofold split due to the spin dependence
of the single particle energy $\varepsilon({\bf p})$ in
Eq.~\p{7.2}. The branches $\omega_{\pm}$ correspond to neutrons
with spin up and spin down.

The distribution functions $f$ should satisfy the norma\-lization
conditions
\begin{align} \frac{2}{\cal
V}\sum_{\bf p}f_{0}({\bf p})&=\varrho,\label{3.1}\\
\frac{2}{\cal V}\sum_{\bf p}f_{3}({\bf
p})&=\varrho_\uparrow-\varrho_\downarrow\equiv\Delta\varrho.\label{3.2}
 \end{align}
 Here $\varrho=\varrho_{\uparrow}+\varrho_{\downarrow}$ is the total density of
 neutron matter, $\varrho_{\uparrow}$ and $\varrho_{\downarrow}$  are the neutron number densities
 with spin up and spin down,
 respectively. The
quantity $\Delta\varrho$  may be regarded as the neutron spin
order parameter. It determines the magnetization of the system
$M=\mu_n \Delta\varrho$, $\mu_n$ being the neutron magnetic
moment. The magnetization may contribute to the internal magnetic
field $B=H+4\pi M$. However, we will assume, analogously to
Refs.~\cite{PG,BPL}, that the contribution of the magnetization
 to the magnetic field
$B$ remains small for all relevant densities and magnetic field
strengths, and, hence, \bal B\approx H.\label{approx}\end{align}
This assumption holds true due to the tiny value of the neutron
magnetic moment
$\mu_n=-1.9130427(5)\mu_N\approx-6.031\cdot10^{-18}$
MeV/G~\cite{A} ($\mu_N$ being the nuclear magneton)
 and is confirmed numerically by finding
 solutions of the self-consistent equations in two
approximations, corresponding to preserving and neglecting  the
contribution of the magnetization.

In order to get the self--consistent equations for the components
of the single particle energy, one has to set the energy
functional of the system. In view of the approximation~\p{approx},
it reads~\cite{AIP,IY}
\bal E(f)&=E_0(f,H)+E_{int}(f)+E_{field},\label{enfunc} \\
{E}_0(f,H)&=2\sum\limits_{ \bf p}^{} \varepsilon_0({\bf
p})f_{0}({\bf p})-2\mu_n H\sum\limits_{ \bf p}^{} f_{3}({\bf
p}),\nonumber
\\ {E}_{int}(f)&=\sum\limits_{ \bf p}^{}\{
\tilde\varepsilon_{0}({\bf p})f_{0}({\bf p})+
\tilde\varepsilon_{3}({\bf p})f_{3}({\bf p})\},\nonumber\\
E_{field}&=\frac{H^2}{8\pi}\cal V,\nonumber\end{align} where
\begin{align}\tilde\varepsilon_{0}({\bf p})&=\frac{1}{2\cal
V}\sum_{\bf q}U_0^n({\bf k})f_{0}({\bf
q}),\;{\bf k}=\frac{{\bf p}-{\bf q}}{2}, \label{flenergies}\\
\tilde\varepsilon_{3}({\bf p})&=\frac{1}{2\cal V}\sum_{\bf
q}U_1^n({\bf k})f_{3}({\bf q}).\nonumber
\end{align}
Here  $\varepsilon_0({\bf p})=\frac{{\bf p}^{\,2}}{2m_{0}}$ is the
free single particle spectrum, $m_0$ is the bare mass of a
neutron, $U_0^n({\bf k}), U_1^n({\bf k})$ are the normal Fermi
liquid (FL) amplitudes, and
$\tilde\varepsilon_{0},\tilde\varepsilon_{3}$ are the FL
corrections to the free single particle spectrum. Note that in
this study we will not be interested in the total energy density
and pressure in the interior of a  neutron star. By this reason,
the field contribution $E_{field}$, being the energy of the
magnetic field in the absence of matter, can be omitted. Using
Eqs.~\p{1} and \p{enfunc}, we get the self-consistent equations in
the form \bal\xi_{0}({\bf p})&=\varepsilon_{0}({\bf
p})+\tilde\varepsilon_{0}({\bf p})-\mu_0,\; \xi_{3}({\bf
p})=-\mu_nH+\tilde\varepsilon_{3}({\bf p}).\label{14.2}
\end{align}

   To obtain
 numerical results, we  utilize the  effective Skyrme interaction.
The amplitude of NN interaction for the Skyrme effective forces
reads~\cite{VB} \bal\hat v({\bf p},{\bf
q})&=t_0(1+x_0P_\sigma)+\frac{1}{6}t_3(1+x_3P_\sigma)\varrho^\beta
\label{49}\\&+\frac{1}{2\hbar^2} t_1(1+x_1P_\sigma)({\bf p}^2+{\bf
q}^2) +\frac{t_2}{\hbar^2}(1+x_2P_\sigma){\bf p}{\bf
q},\nonumber\end{align} where
$P_\sigma=(1+{{\boldsymbol\sigma_1\boldsymbol\sigma_2}})/2$ is the
spin exchange operator,  $t_i, x_i$ and $\beta$ are some
phenomenological parameters specifying a given parametrization of
the Skyrme interaction. In Eq.~\p{49}, the spin-orbit term
irrelevant for a uniform matter was omitted. The normal FL
amplitudes can be expressed in terms of the Skyrme
  force parameters~\cite{AIP,IY3}:
\bal U_0^n({\bf k})&=2t_0(1-x_0)+\frac{t_3}{3}\varrho^\beta(1-x_3)\label{101}\\&\quad
+\frac{2}{\hbar^2}[t_1(1-x_1)+3t_2(1+x_2)]{\bf k}^{2},
\nonumber\\
U_1^n({\bf
k})&=-2t_0(1-x_0)-\frac{t_3}{3}\varrho^\beta(1-x_3)\label{102}\\&\quad
+\frac{2}{\hbar^2}[t_2(1+x_2)-t_1(1-x_1)]{\bf k}^{2}\equiv
a_n+b_n{\bf k}^{2}.\nonumber\end{align} Further we do not take
into account the effective tensor forces, which lead to coupling
of the momentum and spin degrees of freedom \cite{HJ,D,FMS}, and,
correspondingly, to anisotropy in the momentum dependence of FL
amplitudes with respect to the spin quantization axis. Then
\begin{align}
\xi_{0}&=\frac{p^2}{2m_{n}}-\mu,\label{4.32}\\
\xi_{3}&=-\mu_nH+(a_n+b_n\frac{{\bf
p}^{2}}{4})\frac{\Delta\varrho}{4}+\frac{b_n}{16}\langle {\bf
q}^{2}\rangle_{3}, \label{4.33}
\end{align}
where the effective neutron mass $m_{n}$  is defined by
 the formula
\bal \frac{\hbar^2}{2m_{n}}=\frac{\hbar^2}{2m_0}+\frac{\varrho}{8}
[t_1(1-x_1)+3t_2(1+x_2)],\label{181}\end{align} and the
renormalized chemical potential $\mu$ should be determined from
Eq.~\p{3.1}. The quantity $\langle {\bf q}^{2}\rangle_{3}$ in
Eq.~\p{4.33}  is the second order moment of the distribution
function $f_3$:
\begin{align} \langle {\bf
q}^{2}\rangle_{3}&=\frac{2}{V}\sum_{\bf q}{\bf q}^2f_{3}({\bf
q}).\label{6.11}\end{align}  In view of Eqs.~\p{4.32}, \p{4.33},
 the branches $\omega_\pm\equiv\omega_\sigma$ of the quasiparticle spectrum
in Eq.~\p{omega} read \beqe
\omega_\sigma=\frac{p^2}{2m_{\sigma}}-\mu+\sigma\bigl(-\mu_nH+\frac{a_n\Delta\varrho}{4}
+\frac{b_n}{16}\langle {\bf
q}^{2}\rangle_{3}\bigr),\label{spectrud}\end{equation} where
$m_\sigma$ is the effective  mass of a neutron with spin up
($\sigma=+1$) and spin down ($\sigma=-1$) \bal
\frac{\hbar^2}{2m_{\sigma}}&=\frac{\hbar^2}{2m_0}
+\frac{\varrho_\sigma}{2}
t_2(1+x_2)\label{187}\\&\quad+\frac{\varrho_{-\sigma}}{4}[t_1(1-x_1)+t_2(1+x_2)],\;
\varrho_{+(-)}\equiv\varrho_{\uparrow(\downarrow)}.\nonumber\end{align}

Note that for totally spin polarized neutron matter \bal
\frac{m_0}{m^*}=1+\frac{\varrho
m_0}{\hbar^2}t_2(1+x_2),\label{masspol}
\end{align}
where $m^*$ is the effective neutron mass in the fully polarized
state. Since usually for Skyrme parametrizations $t_2<0$, we have
the constraint $x_2\leq-1$, which guarantees the stability of
totally polarized neutron matter at high densities.

 It follows from Eq.~\p{spectrud} that the effective chemical
potential $\mu_\sigma$ for neutrons with spin-up ($\sigma=1$) and
spin-down ($\sigma=-1$) can be determined as

\bal \mu_\sigma=\mu+\sigma\bigl(\mu_nH-\frac{a_n\Delta\varrho}{4}
-\frac{b_n}{16}\langle {\bf q}^{2}\rangle_{3}\bigr).\end{align}

 Thus, with account of expressions
\p{2.4}  for the distribution functions $f$, we obtain the
self--consistent equations \p{3.1}, \p{3.2}, and \p{6.11} for the
effective chemical potential $\mu$,
  spin  order parameter
$\Delta\varrho$,
 and  second order moment
$\langle {\bf q}^{2}\rangle_{3}$.

\section{Solutions of self-consistent equations at $T=0$. Thermodynamic stability}

Here we directly solve the self-consistent equations at zero
temperature and present  the neutron spin order parameter as a
function of density and magnetic field strength. In solving
numerically  the self-consistent equations, we utilize SLy4 and
SLy7 Skyrme forces~\cite{CBH}, which were constrained originally
to reproduce the results of microscopic neutron matter
calculations (pressure versus density curve). Note that the
density dependence of the nuclear symmetry energy, calculated with
these Skyrme interactions, gives the neutron star models in a
broad agreement with the observables such as the minimum rotation
period, gravitational mass-radius relation, the binding energy,
released in supernova collapse, etc.~\cite{RMK}. Besides, these
Skyrme parametrizations satisfy the constraint $x_2\leq-1$,
obtained from Eq.~\p{masspol}.

We consider magnetic fields up to the values allowed by the scalar
virial theorem. For a neutron star with the mass $M$ and radius
$R$, equating the magnetic field energy $E_H\sim (4\pi
R^3/3)(H^2/8\pi)$ with the gravitational binding energy $E_G\sim
GM^2/R$, one gets the estimate
$H_{max}\sim\frac{M}{R^2}(6G)^{1/2}$. For a typical neutron star
with $M=1.5M_{\odot}$ and $R=10^{-5}R_\odot$, this yields for the
maximum value of the magnetic field strength
$H_{max}\sim10^{18}$~G. This magnitude can be expected in the
interior of a magnetar while recent observations report the
surface values up to $H\sim 10^{15}$~G, as inferred from the
hydrogen spectral lines~\cite{IShS}.

In order to characterize spin ordering in neutron matter, it is
convenient to introduce a neutron spin polarization parameter
\beqe
\Pi=\frac{\varrho_{\uparrow}-\varrho_{\downarrow}}{\varrho}\equiv\frac{\Delta\varrho}{\varrho}.
\end{equation}

\begin{figure}[tb]
\begin{center}
\includegraphics[width=8.6cm,keepaspectratio]{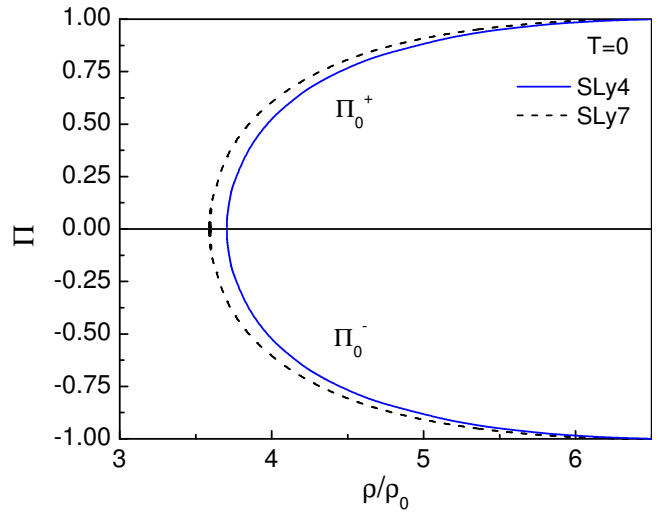}
\end{center}
\vspace{-2ex} \caption{(Color online) Neutron spin polarization
parameter as a function of density at vanishing temperature and
magnetic field.} \label{fig1}\vspace{-0ex}
\end{figure}

Fig.~\ref{fig1} shows the dependence of the neutron spin
polarization parameter from density, normalized to the nuclear
saturation density $\varrho_0$, at zero temperature in the absence
of the magnetic field. The spontaneous polarization develops at
$\varrho=3.70\varrho_0$ for the SLy4 interaction ($\varrho_0=0.16\
\rm{fm}^{-3}$) and at $\varrho=3.59\varrho_0$ for the SLy7
interaction ($\varrho_0=0.158\ \rm{fm}^{-3}$), that reflects the
instability of neutron matter with the Skyrme interaction at such
densities against  spin fluctuations. Since
the self-consistent equations at $H=0$ 
 are
invariant with respect to the global flip of neutron spins, we
have two branches  of solutions for the spin polarization
parameter, $\Pi_0^+(\varrho)$ (upper) and $\Pi_0^-(\varrho)$
(lower) which differ only by sign,
$\Pi_0^+(\varrho)=-\Pi_0^-(\varrho)$.

\begin{figure}[tb]
\begin{center}
\includegraphics[width=8.6cm,keepaspectratio]{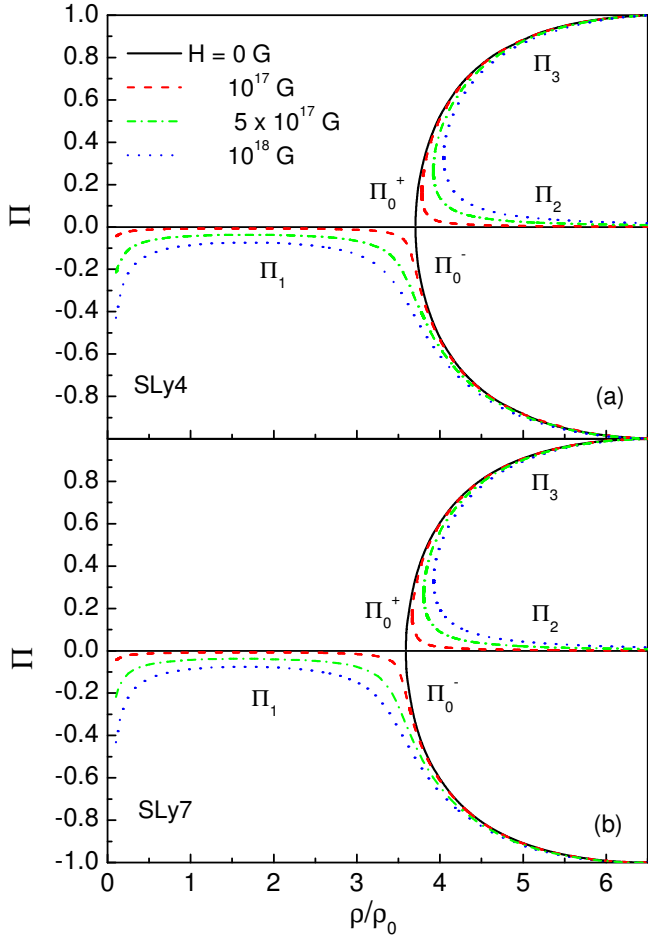}
\end{center}
\vspace{-2ex} \caption{(Color online) Neutron spin polarization
parameter as a function of density at $T=0$ and different magnetic
field strengths for (a) SLy4 interaction and (b) SLy7 interaction.
The branches of spontaneous polarization $\Pi_0^-,\Pi_0^+$ are
shown by solid curves.} \label{fig2}\vspace{-0ex}
\end{figure}
Fig.~\ref{fig2} shows the neutron spin polarization parameter  as
a function of density for a set of fixed values of the magnetic
field. The  branches of spontaneous polarization are modified by
the magnetic field differently,  since the self-consistent
equations at $H\not=0$ lose the invariance with respect to the
global flip of the spins. At nonvanishing $H$, the lower branch
$\Pi_1(\varrho)$, corresponding to the negative spin polarization,
extends down to the very low densities. There are three
characteristic density domains for this branch. At  low densities
$\varrho\lesssim 0.5\varrho_0$, the absolute value of the spin
polarization parameter increases with decreasing density. At
intermediate densities
$0.5\varrho_0\lesssim\varrho\lesssim3\varrho_0$, there is a
plateau in the $\Pi_1(\varrho)$  dependence, whose characteristic
value depends on $H$, e.g., $\Pi_1\approx-0.08$ at $H=10^{18}$~G.
At densities $\varrho\gtrsim3\varrho_0$,  the magnitude of the
spin polarization parameter increases with density, and neutrons
become totally polarized at $\varrho\approx6\varrho_0$.

Note that the results in the low-density domain should be
considered as a first approximation to the real complex picture,
since, as discussed in detail in Ref.~\cite{PG}, the low density
neutron-rich matter in $\beta$-equilibrium  possesses a frustrated
state, "nuclear pasta", arising as a result of competition of
Coulomb long-range interactions and nuclear short-range forces. In
our case, where a pure neutron matter is considered, there is no
mechanical instability due to the absence of the Coulomb
interaction. However, the possibility of appearance of low-density
nuclear magnetic pasta and its impact  on the neutrino opacities
in the protoneutron star early cooling stage should be explored in
a more detailed analysis.

\begin{figure}[tb]
\begin{center}
\includegraphics[width=8.6cm,keepaspectratio]{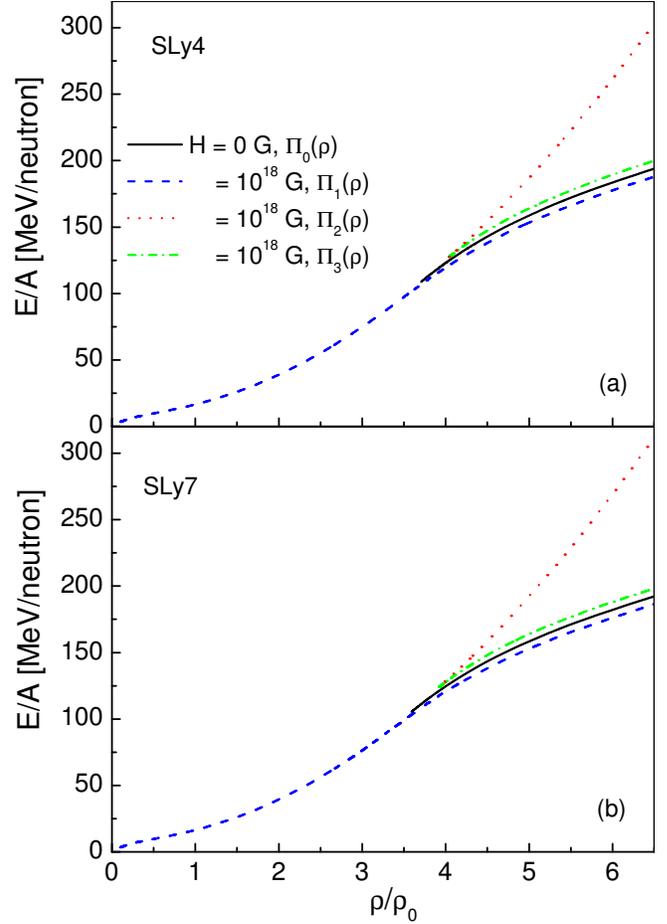}
\end{center}
\vspace{-2ex} \caption{(Color online) Energy  per neutron  as a
function of density at $T=0$  for different branches
$\Pi_1(\varrho)$-$\Pi_3(\varrho)$ of solutions of the
self-consistent equations at $H=10^{18}$~G for (a) SLy4 and (b)
SLy7 interactions, including a spontaneously polarized state.}
\label{fig3}\vspace{-0ex}
\end{figure}

Let us consider the modification of the upper branch of
spontaneous polarization $\Pi_0^+(\varrho)$ at nonvanishing
magnetic field. It is seen from Fig.~\ref{fig2} that now beginning
from some threshold density the self-consistent equations at a
given density have two positive solutions  for the spin
polarization parameter (apart from one negative solution). These
solutions belong to two branches, $\Pi_2(\varrho)$ and
$\Pi_3(\varrho)$, characterized by  different dependence from
density. For the branch $\Pi_2(\varrho)$, the spin polarization
parameter decreases with density and tends to zero value while for
the  branch $\Pi_3(\varrho)$ it increases with density and is
saturated. These branches appear step-wise at the same threshold
density $\varrho_{\rm th}$ dependent on the magnetic field and
being larger than the critical density of spontaneous spin
instability in neutron matter.  For example, for SLy7 interaction,
$\varrho_{\rm th}\approx 3.80\,\varrho_0$ at $H=5\cdot 10^{17}$~G,
and  $\varrho_{\rm th}\approx 3.92\,\varrho_0$ at $H= 10^{18}$~G.
The magnetic field, due to the negative value of the neutron
magnetic moment, tends to orient the neutron spins opposite to the
magnetic field direction. As a result, the spin polarization
parameter for the branches $\Pi_2(\varrho)$, $\Pi_3(\varrho)$ with
the positive spin polarization is smaller than that for the branch
of spontaneous polarization $\Pi_0^+$, and, vice versa, the
magnitude of the spin polarization parameter for the branch
$\Pi_1(\varrho)$ with the negative spin polarization is larger
than the corresponding value for the branch of spontaneous
polarization $\Pi_0^-$. Note that the impact of even such strong
magnetic field as $H=10^{17}$~G is small: The spin polarization
parameter for all three branches $\Pi_1(\varrho)$-$\Pi_3(\varrho)$
is either close to zero, or close to its value in the state with
spontaneous polarization, which  is governed by the spin-dependent
medium correlations.

Thus, at densities larger than $\varrho_{\rm th}$, we have three
branches of solutions: one of them, $\Pi_1(\varrho)$,  with the
negative spin polarization and two others, $\Pi_2(\varrho)$ and
$\Pi_3(\varrho)$, with the positive polarization. In order to
clarify, which branch is thermodynamically preferable, we should
compare the corresponding free energies. Fig.~\ref{fig3} shows the
energy per neutron as a function of density at $T=0$ and
$H=10^{18}$~G for these three branches, compared with the energy
per neutron for a spontaneously polarized state [the branches
$\Pi_0^\pm(\varrho)$]. It is seen that the state with the majority
of neutron spins  oriented opposite to the direction of the
magnetic field [the branch $\Pi_1(\varrho)$] has a lowest energy.
This result is intuitively clear, since magnetic field tends to
direct the neutron spins opposite to $\bf{H}$, as mentioned
earlier. However, the state, described by the branch
$\Pi_3(\varrho)$ with the positive spin polarization, has the
energy very close to that of the thermodynamically stable state.
This means that despite the presence of a strong magnetic field
$H\sim 10^{18}$~G, the state with the majority of neutron spins
directed  along the magnetic field can be realized as a metastable
state in the dense core of a neutron star in the model
consideration with the Skyrme effective interaction. In this
scenario, since such states exist only at densities
$\varrho\geqslant\varrho_{\rm th}$,
 under decreasing density (going from the
interior to the outer regions of a magnetar) a metastable state
with the positive spin polarization  at the threshold density
$\varrho_{\rm th}$ changes into a thermodynamically stable state
with the negative spin polarization.

\begin{figure}[tb]
\begin{center}
\includegraphics[width=8.6cm,keepaspectratio]{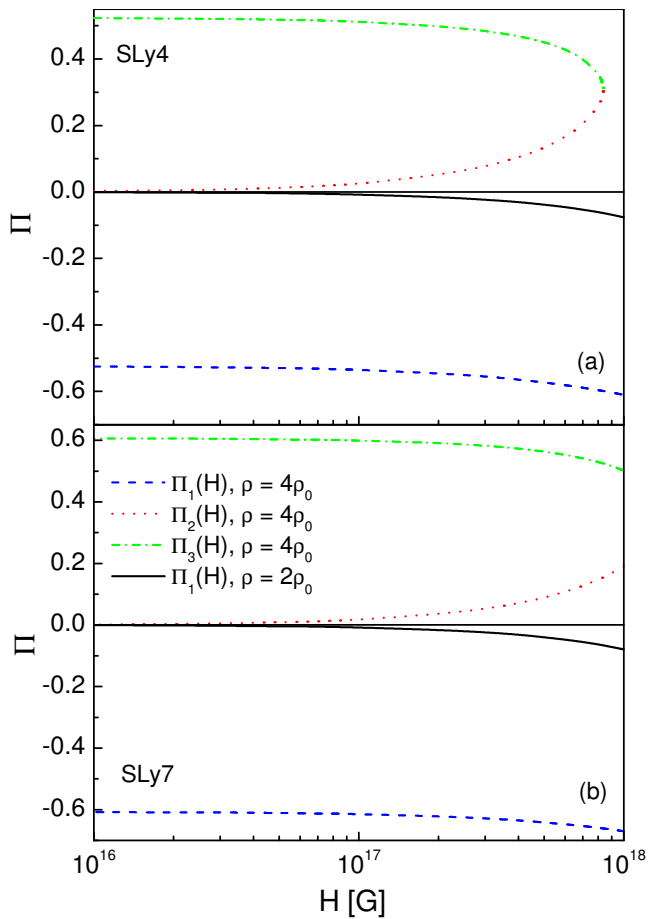}
\end{center}
\vspace{-2ex} \caption{(Color online) Spin polarization parameter
as  a function of the magnetic field strength at $T=0$ for
different branches $\Pi_1(H)$-$\Pi_3(H)$ of solutions of the
self-consistent equations at $\varrho=4\varrho_0$ and for the
branch $\Pi_1(H)$ at $\varrho=2\varrho_0$ for (a) SLy4 interaction
and (b) SLy7  interaction.} \label{fig4}\vspace{-0ex}
\end{figure}

At this point, note some important differences between the results
in our study and those obtained in Ref.~\cite{PG}. First, in the
study~\cite{PG} of neutron matter at a strong magnetic field only
one branch of solutions for the spin polarization parameter was
found in the model with the Skyrme interaction (for the same SLy4
and SLy7 parametrizations). However, in fact, we have seen that
the degenerate branches of spontaneous polarization (at zero
magnetic field) with the positive and negative spin polarization
  are modified differently by the
magnetic field, and, as a result,  in the Skyrme model, in
general, there are three different branches of solutions of the
self-consistent equations at nonvanishing magnetic field. Besides,
the only branch considered in Ref.~\cite{PG} and  corresponding to
our thermodynamically stable branch $\Pi_1$, is characterized by
the positive spin polarization, contrary to our result with
$\Pi_1<0$. This disagreement is explained by the incorrect sign
before the term with the magnetic field in the equation for the
quasiparticle spectrum in Ref.~\cite{PG} (analogous to
Eq.~\p{spectrud} in our case). Clearly,  in the equilibrium
configuration the majority of neutron spins are aligned opposite
to the magnetic field.

\begin{figure}[tb]
\begin{center}
\includegraphics[width=8.6cm,keepaspectratio]{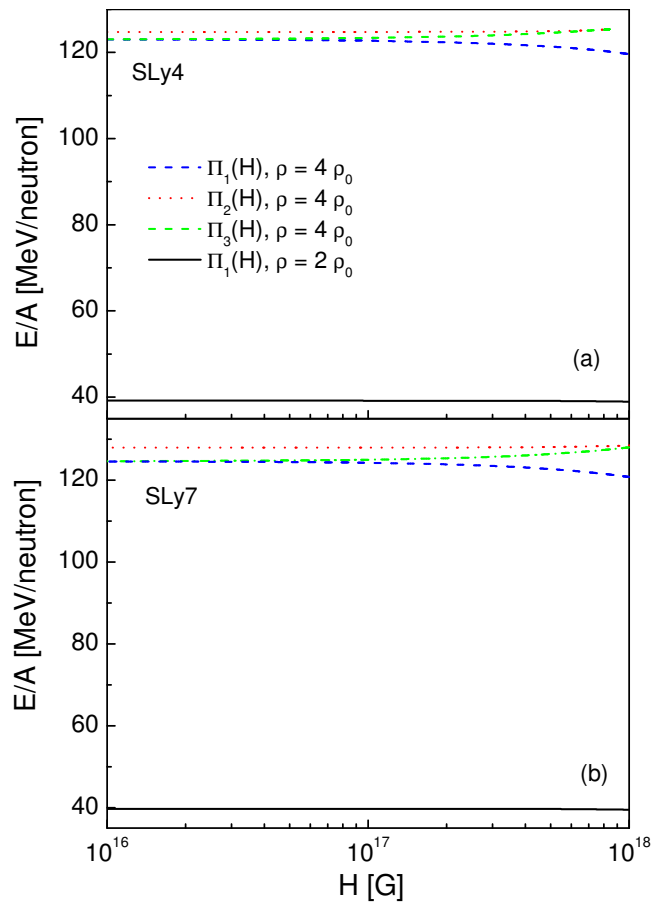}
\end{center}
\vspace{-0ex} \caption{(Color online) Same as in Fig.~\ref{fig4}
but for the energy per neutron.} \label{fig5}
\end{figure}

Fig.~\ref{fig4} shows the spin polarization parameter
  as  a function of the
magnetic field strength at zero temperature  for different
branches $\Pi_1(H)$-$\Pi_3(H)$ of solutions of the self-consistent
equations at $\varrho=4\varrho_0$ compared with that for the
branch $\Pi_1(H)$ at $\varrho=2\varrho_0$. It is seen that up to
the field strengths $H=10^{17}$~G, the influence of the magnetic
field is rather marginal. For the branches $\Pi_1(H)$ and
$\Pi_2(H)$, the magnitude of the spin polarization parameter
increases with the field strength while for the $\Pi_3(H)$ it
decreases. Interestingly, as is clearly seen from the top panel
for the SLy4 interaction, at the given density, there exists some
maximum magnetic field strength $H_m$ at which the branches
$\Pi_2$ and $\Pi_3$ converge and do not continue at $H>H_m$.

Fig.~\ref{fig5} shows the energy of neutron matter per particle as
a function of the magnetic field strength at $T=0$ under the same
assumptions as in Fig.~\ref{fig4}.  It is seen that the state with
the negative spin polarization [branch $\Pi_1(H)$] becomes more
preferable with increasing the magnetic field although the total
effect of changing the magnetic field strength by two orders of
magnitude on  the energy corresponding to all three branches
$\Pi_1(H)$-$\Pi_3(H)$ remains small. It is also seen that the
increase of the density by a factor of two leads to the increase
in the energy per neutron roughly  by a factor of three reflecting
the fact that the medium correlations play more important role in
building the energetics of the system than the impact of a strong
magnetic field.

\section{Conclusions}
We have considered spin polarized states  in neutron matter at a
strong magnetic field in the model with the Skyrme effective NN
interaction (SLy4, SLy7 parametrizations). The self-consistent
equations for the spin polarization parameter and chemical
potential of neutrons have been obtained and analyzed at
 zero temperature. It has been shown that the thermodynamically
stable branch of solutions for the spin polarization parameter as
a function of density corresponds to the case when the majority of
neutron spins are oriented opposite to the direction of the
magnetic field (negative spin polarization). This branch extends
from the very low densities to the high density region where the
spin polarization parameter is saturated, and, respectively,
neutrons become totally spin polarized. Besides, beginning from
some threshold density $\varrho_{\rm th}$ being dependent on the
magnetic field strength the self-consistent equations have also
two other branches (upper and lower) of solutions for the spin
polarization parameter corresponding to the case when the majority
of neutron spins are oriented along the magnetic field (positive
spin polarization). For example, for SLy7 interaction,
$\varrho_{\rm th}\approx 3.80\,\varrho_0$ at $H=5\cdot 10^{17}$~G,
and $\varrho_{\rm th}\approx 3.92\,\varrho_0$ at $H= 10^{18}$~G.
The spin polarization parameter along the upper branch increases
with density and is saturated, while along the lower branch  it
decreases and vanishes. The free energy corresponding to the upper
branch turns out to be very close to the free energy corresponding
to the thermodynamically preferable branch with the negative spin
polarization. As a consequence, at a strong magnetic field, the
state with the positive spin polarization can be realized as a
metastable state at the high density region in neutron matter
which under decreasing density (going from the interior to the
outer regions of a magnetar) at the threshold density
$\varrho_{\rm th}$ changes into a thermodynamically stable state
with the negative spin polarization.

In this study, we have considered the zero temperature case, but
as was shown in Ref.~\cite{PG}, the influence of finite
temperatures on spin polarization remains moderate in the Skyrme
model, at least, up to the temperatures relevant for protoneutron
stars, and, hence, one can expect that the considered scenario
will be preserved at finite temperatures as well.   The possible
existence of a metastable state with positive spin polarization
will affect the neutrino opacities of a neutron star matter in a
strong magnetic field, and, hence, will lead to the change of
cooling rates of a neutron star compared to cooling rates in the
scenario with the majority of neutron spins oriented opposite to
the magnetic field~\cite{PG2}.

The calculations of the neutron spin polarization parameter and
energy per neutron show that the influence of the magnetic field
remains small at the field strengths up to $10^{17}$~G. Note that
in Ref.~\cite{PG} the consideration also has been done for the
Gogny effective  NN interaction (D1S, D1P parametrizations) up to
densities $4\varrho_0$. Since for the D1S parametrization there is
no spontaneous  ferromagnetic transition in neutron matter for all
relevant densities, and for the  D1P parametrization this
transition occurs at the density  larger than
$7\varrho_0$~\cite{LVRP}, no sign of a ferromagnetic transition at
a strong magnetic field was found in Ref.~\cite{PG} up to
densities  $4\varrho_0$ for these Gogny forces. According to our
consideration, one can expect that the metastable states with the
positive spin polarization in neutron matter   at a strong
magnetic field could appear at densities larger than $7\varrho_0$
for the D1P parametrization while the scenario with the only
branch of solutions corresponding to the negative spin
polarization would be realized for the D1S force.

It is worthy to note also that in the given research a neutron
star matter was approximated by pure neutron matter. This
approximation allows one to get the qualitative description of the
spin polarization phenomena  and should be considered as a first
step towards a more realistic description of neutron stars taking
into account a finite fraction of protons with the charge
neutrality and beta equilibrium conditions. In particular, some
admixture of protons can affect the onset densities of enhanced
polarization in a neutron star matter with the Skyrme interaction.

\section*{ACKNOWLEDGEMENTS}

J.Y. was supported  by  grant R32-2008-000-10130-0 from WCU
project of MEST and NRF through Ewha Womans University.

\end{document}